\input{aipcheck}

\documentclass[
    ,final            
  ,numberedheadings 
  ]
  {aipproc}

\layoutstyle{6x9}

\usepackage{amsmath,amssymb,graphicx,mathrsfs}
\usepackage{bm}
\newcommand{\nn}{\nonumber\\}

\newcommand{\f}{\bar{f}}
\newcommand{\F}{\bar{F}}
\newcommand{\V}{\bar{V}}
\newcommand{\T}{\bar{T}}

\newcommand{\ben}{\begin{displaymath}}
\newcommand{\een}{\end{displaymath}}
\newcommand{\be}{\begin{equation}}
\newcommand{\ee}{\end{equation}}
\newcommand{\ba}{\begin{align}}
\newcommand{\ea}{\end{align}}
\newcommand{\bea}{\begin{eqnarray}}
\newcommand{\eea}{\end{eqnarray}}
\newcommand{\eqn}[1]{\label{#1}}
\newcommand{\eq}[1]{Eq.~(\ref{#1})}
\newcommand{\eqs}[1]{Eqs.\ (\ref{#1})}
\newcommand{\fign}[1]{\label{#1}}
\newcommand{\fig}[1]{Fig.\ \ref{#1}}

\begin{document}

\title{Crossing symmetric potential model of pion-nucleon scattering}


\author{B. Blankleider}{
  address={School of Chemical and Physical Sciences, Flinders University, South Australia}
}

\author{A. N. Kvinikhidze}{
  address={Razmadze Mathematical Institute, Republic of Georgia}
}

\author{T. Skawronski}{
  address={School of Chemical and Physical Sciences, Flinders University, South Australia}
}

\begin{abstract}
A crossing symmetric $\pi N$ scattering amplitude is constructed through a complete attachment of two external pions to the dressed nucleon propagator of an underlying $\pi N$ potential model. Our formulation automatically provides expressions also for the crossing symmetric and gauge invariant pion photoproduction and Compton scattering amplitudes. We show that our amplitudes are unitary if they coincide on-shell with the amplitudes obtained by attaching one pion to the dressed $\pi NN$ vertex of the same potential model. 


\end{abstract}

\maketitle


\section{Introduction}

The $\pi N$ scattering amplitude $t$ is often described using the set of equations \cite{Morioka:1981vb,Gross:1992tj,Pascalutsa:2000bs}
\begin{subequations}  \eqn{t} 
\begin{align}
t  &= f g \f + t_b,  &  t_b &= v + t_b G_0 v  , \eqn{ta} \\
\f &= \f_0 + \f_0 G_0 t_b,  & f &= f_0 + t_b G_0 f_0,\eqn{tb} \\
 g &= g_0 + g_0\Sigma g, & \Sigma &= \f_0 G_0 f, \eqn{tc}
\end{align}
\end{subequations}
where $f$ ($f_0$) is the dressed (bare)  $N\rightarrow \pi N$ vertex, $\f$ ($\f_0$) is the dressed (bare)  $\pi N\rightarrow  N$ vertex, $g$ ($g_0$) is the dressed (bare) nucleon propagator, $G_0$ is the renormalized disconnected $\pi N$ propagator, and  $t_b$ ($v$) is the  "non-pole" $\pi N$ t-matrix (potential) with the pole term $fg\f$ ($f_0g_0\f_0$) removed. Although these equations  provide an exact description in full field theory, their main feature is that they allow one to preserve unitarity when making models for the potential $v$ and bare vertex $f_0$. However, like all potential models, these equations suffer from a lack of crossing symmetry, a property  whose importance has  been emphasized for more than 50 years \cite{Chew:1955zz,McLeod:1984zz,FernandezRamirez:2008pz}.

Similarly, the pion photoproduction amplitude $t^\gamma$  is often descsribed by a set of equations that essentially result from \eq{ta} and \eq{tb} by replacing the initial pion with a photon
\cite{Nozawa:1989pu,Surya:1995ur,Pascalutsa:2004pk}:
\begin{subequations} \eqn{M} 
\begin{align}
t^\gamma &= f g \f^\gamma + t_b^\gamma,  &  t_b^\gamma &= v^\gamma + t_b G_0 v^\gamma  ,\\
\f^\gamma &= \f^\gamma_0 + \f_0 G_0 t_b^\gamma,   & 
\end{align}
\end{subequations}
where $\f^\gamma$ ($\f^\gamma_0$) is the dressed (bare)  $\gamma N\rightarrow  N$  vertex, and  $t^\gamma_b$ ($v^\gamma$) is the pion photoproduction amplitude (Born term) with the pole term $fg\f^\gamma$ ($f_0g_0\f^\gamma_0$) removed. Once again the feature of these equations is that they respect unitarity. This time, however, these equations suffer not only from a lack of crossing symmetry, but also from the breaking of manifest gauge invariance (because the photon is not coupled to all places in the underlying field theory). We shall refer to \eqs{t} and \eqs{M} as the {\em standard} description.

In this paper we present new equations for $\pi N$ scattering, pion photoproduction, and Compton scattering, that are based on the potential model of \eqs{t}, but that preserve crossing symmetry and manifest gauge invariance. Our approach is based on the idea of coupling external pions and photons to all possible places in the dressed propagator $g$ of \eq{tc}, and is achieved using the gauging of equations method \cite{KB3,Kvinikhidze:1999xp}. Like the standard description, our approach is exact in full field theory;\footnote{For simplicity of presentation, we ignore any terms that cannot be obtained by the attachment of external pions or photons.  Such contributions, if present,  are gauge invariant and crossing symmetric on their own, and can therefore  be separately added to our derived amplitudes.} 
 however, just opposite to the standard description, when models are made for the potential $v$ and bare vertex $f_0$, our approach preserves crossing symmetry and gauge invariance at the expense of  unitarity. The lack of built-in unitarity is not surprising since our approach effectively sums the full perturbation  series in a way that is different from the usual method of iterating a kernel. 
Nevertheless, we show that our amplitudes will satisfy unitarity  whenever the crossing symmetric $\pi N$ amplitude coincides, on-shell, with the one obtained by attaching one pion to the dressed $\pi NN$ vertex $f$ of \eq{tb}.

\section{Single-gauged amplitude}
In Refs.\ \cite{KB3,Kvinikhidze:1999xp} we introduced a technique for attaching an external photon to all possible places (vertices, propagators, potentials, etc.) within a strongly interacting system described by dynamical equations. The completeness of the attachment led to the gauge invariance of the resulting electromagnetic currents. Here we use the same technique to also attach first one external pion, and then in the next section, a second external pion, in order to achieve our goal of deriving a crossing symmetric $\pi N$ scattering  amplitude. 

We begin by applying our gauging technique to the $\pi N$ Green function $G$ generated by the non-pole potential $v$:
\be 
G = G_0 + G_0 v G.  \eqn{G}
\ee
Denoting by $G^\mu$ the 5-point function resulting from a complete attachment of an external pion to $G$, \eq{G} is "gauged"\footnote{We shall use "gauging" to mean the process of attaching any external particle, not just a gauge boson.} to obtain
\be
G^\mu = G^\mu_0 + G^\mu_0 v G+G_0 v^\mu G+G_0 v G^\mu
\ee
which is easily solved to get
\begin{align}
G^\mu &= G \Lambda^\mu G, &  \Lambda^\mu &= \Lambda_0^\mu + v^\mu
\end{align}
where $\Lambda_0^\mu \equiv G_0^{-1} G_0^\mu G_0^{-1}$  is a vertex function derived by attaching an external pion to the disconnected $\pi N$ propagator $G_0$. Note that $G_0=g_\pi g_N$ where $g_\pi$ is the pion propagator and $g_N$ is the renormalized nucleon propagator (in exact field theory $g_N=g/Z$ where $Z$ is the renormalization constant; however, in model calculations one usually takes $g_N$ to be the bare propagator with physical nucleon mass).
As G-parity conservation forbids a three-pion vertex,  $\Lambda_0^\mu = \Gamma_N^\mu g_\pi^{-1}$ where $\Gamma_N^\mu\equiv g_N^{-1}g_N^\mu g_N^{-1}$. Thus
\be
\Lambda^\mu =  \Gamma_N^\mu g_\pi^{-1}+v^\mu.
\ee
The gauged potential $v^\mu$ can be constructed phenomenologically, or derived by gauging a  specific model for $v$. Similarly, the gauging of the dressed nucleon propagator $g$ of \eq{tc}, gives the 3-point function $g^\mu$:
\begin{align}
g^\mu &= g \Gamma^\mu g, &  \Gamma^\mu &= \Gamma_0^\mu + \Sigma^\mu \eqn{gmu}
\end{align}
where $\Gamma^\mu$  is the dressed $\pi NN$ vertex function, 
$\Gamma_0^\mu \equiv g_0^{-1} g_0^\mu g_0^{-1}$   is the bare vertex function, and  $\Sigma^\mu$ is the gauged dressing. Gauging $\Sigma = \f_0 G f_0$ then leads to a simple intuitive expression for the  $\pi NN$ dressed  vertex function:
\begin{align}
\Gamma^\mu 
 &= \Gamma_0^\mu + \f^\mu_0 G_0 f + \f G_0 f^\mu_0 + \f G_0^\mu f
+\f G_0 v^\mu G_0 f , \eqn{Gammamu}
\end{align}
which is illustrated in Fig.\ (1).
\begin{figure}[t]
  \includegraphics[width=11cm]{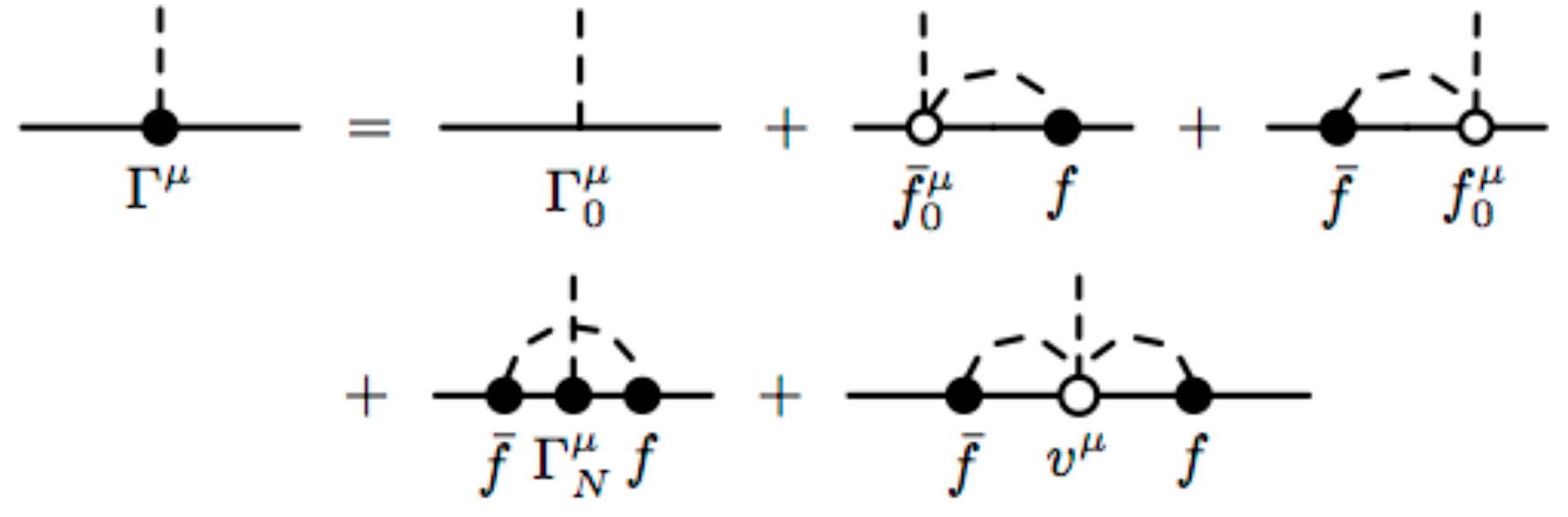}
  \caption{The dressed $\pi NN$ vertex function $\Gamma^\mu$ resulting from a complete pion attachment to the dressed nucleon propagator $g$ defined by \eq{tc}.}
\end{figure}

The main results of this section come from the gauging of the dressed $\pi NN$ vertrex $f$:
\begin{align}
(G_0 f g)^\nu &= (Gf_0g)^\nu = G^\nu f_0 g + G f_0^\nu g + G f_0 g^\nu\nn
&= G\Lambda^\nu G_0 f g + G f_0^\nu g + G_0 f g \Gamma^\nu g.
\end{align}
Cutting off the external legs immediately gives the amplitude $T^\nu\equiv G_0^{-1}(G_0 f g)^\nu g^{-1}$:
\be
T^\nu =  (1+ t_b G_0)(\Lambda_0^\nu G_0 f + v^\mu G_0 f+f_0^\nu) + f g \Gamma^\nu. \eqn{Tnu}
\ee
If superscript $\nu$ corresponds to an external photon, then $T^\nu \sqrt{Z}$ is the properly normalized manifestly gauge invariant pion photoproduction amplitude, originally derived in Ref.\ \cite{vanAntwerpen:1994vh} using a more involved approach. If superscript $\nu$ corresponds to an external pion, then $T^\nu \sqrt{Z}$ is a "hybrid" $\pi N$ scattering amplitude where the {\em initial} state pion is due to gauging and the final state pion is due to the original standard description. In a similar way, vertex $\f$ can be gauged to obtain the hybrid amplitude $\T^\mu$ where the {\em final} 
state pion is due to gauging. One can express these amplitudes in a "pole plus non-pole" form analogous to \eq{ta}:
\begin{subequations}\eqn{T}
\begin{align} 
T^\nu &=  f g \Gamma^\nu + T_b^\nu, & T^\nu_b &= (1+t_bG_0)V^\nu, & V^\nu &= f_0^\nu +\Lambda^\nu G_0 f,\\
\T^\mu &=  \Gamma^\mu g \f+ \T_b^\mu, & \T^\mu_b &= \V^\mu(1+G_0 t_b), &  \V^\mu &= \f_0^\mu +\f  G_0 \Lambda^\mu.
\end{align}
\end{subequations}
It follows that amplitudes $T_b^\nu$ and $\T_b^\mu$ satisfy non-standard Bethe-Salpeter equations
\begin{align}
T_b^\nu &= V^\nu + v G_0 T_b^\nu, & \T_b^\mu &= \V^\mu + \T_b^\mu  G_0 v \eqn{TBS}
\end{align}
where the kernel and inhomogeneous terms involve $\pi N$ potentials of different origin. The amplitudes
are clearly not crossing symmetric.

We can now express \eq{Gammamu} in the following two ways analogous to \eq{tb}:
\begin{subequations} \eqn{F}
\begin{align}
\Gamma^\nu &= \F_0^\nu + \f G_0 V^\nu , & \F_0^\nu &= \Gamma_0^\nu +  \f^\nu_0 G_0 f  
\eqn{Fa}, \\
\Gamma^\mu &= F_0^\mu + \V^\mu G_0 f, & F_0^\mu &= \Gamma_0^\mu + \f G_0 f^\mu_0 ,
\eqn{Fb} 
\end{align}
\end{subequations}
where \eq{Fa} and \eq{Fb} are to be used to describe pion (or photon) absorption and creation  vertices, respectively.
In exact field theory, $\Gamma^\mu$ can be identified with function $f/\sqrt{Z}$ of the standard description. \eqs{F} then imply the following identities (in exact field theory) relating the standard description quantities $f_0$ and $v$ (typically the model inputs) to their gauged counterparts:
\begin{align}
f_0/\sqrt{Z} &= F_0^\mu,  &    v/\sqrt{Z} &= \V^\mu ; &
\f_0/\sqrt{Z} &= \F_0^\nu,  &    v/\sqrt{Z} &= V^\nu . \eqn{key}
\end{align}
\begin{figure}[t] \fign{fig:key}
  \includegraphics[width=12cm]{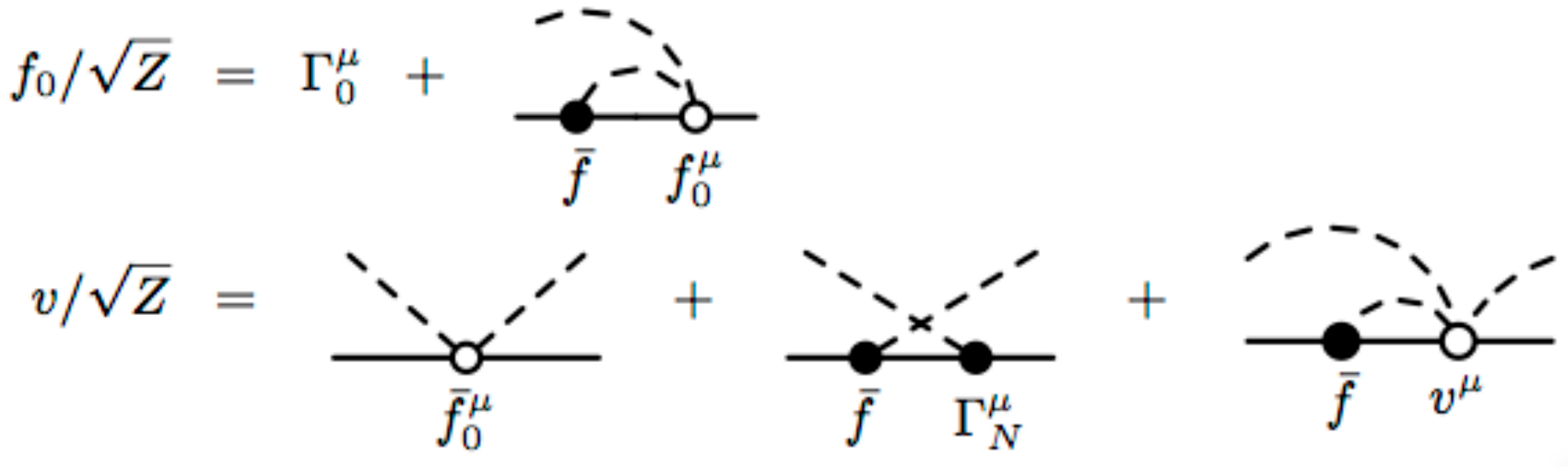}
  \caption{In exact field theory, relations connecting the bare $\pi NN$ vertex $f_0$ and "non-pole" $\pi N$  potential $v$ of the standard description, \eqs{t}, to the corresponding gauged quantities $f_0^\mu$ and $v^\mu$ of the single-gauged description - see the first two of \eqs{key}.}
\end{figure}
The first two of \eqs{key} are illustrated in \fig{fig:key}.

\subsection{Unitarity}
For ease of presentation, we discuss unitarity within the framework of time ordered perturbation theory for which 
\be
G_0(E^+) - G_0(E^-) = -2\pi i \delta(E-H_0)
\ee
where $H_0$ is the free Hamiltonian and $E^\pm=E\pm i\epsilon$.
We consider only 2-body unitarity and thus restrict the discussion to energies $E$ below the two-pion threshold. In this energy region the potential $v$ and bare vertex $f_0$ are real, and the standard description of \eqs{t} will therefore satisfy the following unitarity relations:
 \begin{subequations} \eqn{U}
\begin{align}
t-t^\dagger &=  t^\dagger \!  \delta \, t,  & t_b-t_b^\dagger &=  t_b^\dagger \!  \delta \, t_b,\\
f-f^\dagger &=  t_b^\dagger \delta f, &  \f-\f^\dagger &= \f^\dagger \delta t_b,\\
g-g^\dagger &=   g^\dagger \, (\Sigma-\Sigma^\dagger) \, g , &
\Sigma-\Sigma^\dagger &=  \f{\,}^\dagger   \delta \, f,\\
G-G^\dagger &= (1+G_0^\dagger\, t_b^\dagger) \,  \delta \, (1+t_bG_0),  & &  
\eqn{unit}
\end{align}
 \end{subequations}
where $\delta$ is shorthand for $-2\pi i \delta(E-H_0)$ and  
 where identical quantities with and without a dagger represent the same functions of $E^-$ and $E^+$, respectively (i.e., a dagger does not mean Hermitean conjugate, but rather, $T \equiv T(E^+)$ and $T^\dagger \equiv T(E^-)$). Applying these relations to \eqs{T} and \eqs{F} one obtains the analogous unitarity relations for the hybrid amplitudes:
\begin{subequations}  \eqn{UH}
\begin{align}
T^\nu-T^\nu{}^\dagger &=  t^\dagger \!  \delta \, T^\nu,  
& T^\nu_b-T_b^\nu{}^\dagger &=  t_b^\dagger \!  \delta \, T^\nu_b,   \eqn{UHa} \\
\T^\mu-\T^\mu{}^\dagger &=  \T^\mu{}^\dagger \!  \delta \, t,  
& \T^\mu_b-\T_b^\mu{}^\dagger &=  \T_b^\mu{}^\dagger \!  \delta \, t_b,\\
\Gamma^\mu-\Gamma^\mu{}^\dagger &= \T_b^\mu{}^\dagger \delta f, &
\Gamma^\nu-\Gamma^\nu{}^\dagger &= \f^\dagger \delta T_b^\nu .
\end{align}
 \end{subequations}
In the case of gauging with photons, \eq{UHa} provides just the usual statement of Watson's theorem for the gauge invariant pion photoproduction amplitude $T^\nu$. However, in the case of gauging with pions, \eq{UHa} differs from the  usual statement of unitarity for the hybrid amplitude in that a $T^\nu{}^\dagger$ has been replaced with a $t^\dagger$ from the standard description. Thus the only way for a hybrid $\pi N$ amplitude $T^\nu$ to be unitary is for it to coincide, on-shell, with the  standard amplitude $t$.

\section{Double-gauged amplitude}

To obtain a crossing symmetric $\pi N$ scattering amplitude, we attach two external pions to the dressed nucleon propagator $g$; that is, we first gauge $g$ to obtain $g^\mu$ as in \eq{gmu}, and then gauge \eq{gmu} to obtain $g^{\mu\nu}$ with indices $\nu$ and $\mu$ denoting the initial- and final-state pion, respectively. Thus
\begin{align}
g^{\mu\nu} &= gT^{\mu\nu} g,  &
T^{\mu\nu} &=\Gamma^\mu g \Gamma^\nu+\Gamma^\nu g\Gamma^\mu + \Gamma_0^{\mu\nu} +\Sigma^{\mu\nu}  \eqn{gmunu}
\end{align}
where $ZT^{\mu\nu}$ is the properly normalized crossing symmetric $\pi N$ amplitude. In \eq{gmunu}
\begin{subequations}
\begin{align}
 \Gamma_0^{\mu\nu} &= \left(g_0^{-1} g_0^\mu g_0^{-1}\right)^\nu 
 =g_0^{-1} g_0^{\mu\nu} g_0^{-1} - \Gamma_0^\mu g_0 \Gamma_0^\nu  - \Gamma_0^\nu g_0 \Gamma_0^\mu ,\\[2mm]
\Sigma^{\mu\nu}
 &= \f_0^{\mu\nu} G_0 f + \f G_0 f_0^{\mu\nu} + \f G_0  \Lambda^{\mu\nu} G_0 f 
 +A^{\mu\nu}+A^{\nu\mu}
 \end{align}
 \end{subequations}
 where
 \begin{subequations}
 \begin{align}
 \Lambda^{\mu\nu} &=\Gamma_N^{\mu\nu} g_\pi^{-1}+ v^{\mu\nu}, \\
A^{\mu\nu} 
&= \V^\mu G V^\nu = \V^\mu G_0 T_b^\nu = \T_b^\mu G _0V^\nu . \eqn{A}
\end{align}
\end{subequations}
One can thus write the crossing symmetric  $ T^{\mu\nu}$ in terms of pole and non-pole parts as
 \begin{subequations} \eqn{TVmunu}
 \begin{align}
 T^{\mu\nu} &= \Gamma^\mu g \Gamma^\nu +  T_b^{\mu\nu}, \hspace{2cm}
 T_b^{\mu\nu}  =  V^{\mu\nu} +\T_b^\mu G _0V^\nu, \eqn{Tmunu} \\[1mm]
V^{\mu\nu} &=  \Gamma^\nu g\Gamma^\mu + \Gamma_0^{\mu\nu} 
+\f_0^{\mu\nu} G_0 f + \f G_0 f_0^{\mu\nu} + \f G_0  \Lambda^{\mu\nu} G_0 f 
 +A^{\nu\mu}   \nn
 &=\left(\F_0^\nu\right)^\mu + \Gamma^\nu g \Gamma^\mu + \f G_0 \left(V^\nu\right)^\mu .
  \eqn{Vmunu}
 \end{align}
 \end{subequations}
We note that the second of \eqs{Tmunu} still has the basic structure of a Bethe-Salpeter equation although neither the two non-pole $\pi N$ potentials $V^{\mu\nu}$ and $V^\nu$, nor the two non-pole t matrices $T_b^{\mu\nu}$ and $\T_b^\mu$ are of the same origin. It is also important to note that \eqs{TVmunu} apply also to the cases where either one or both of the superscripts $\mu$ and $\nu$ refer to the gauging by photons. That is, these equations provide a unified, crossing symmetric description of pion-nucleon elastic scattering, pion photoproduction, and Compton scattering. Morever, the electromagnetic amplitudes are due to a complete attachment of photons and are therefore manifestly gauge invariant. 

\subsection{Unitarity}
 
 In the crossing symmetric formulation, the $\pi N$ potential $V^{\mu\nu}$ of \eq{Vmunu} is real below two-pion threshold, as is the hybrid potential $V^\nu$. It is thus straightforward to obtain the 2-body unitarity relations from \eq{Tmunu} and the unitarity relations for the hybrid amplitudes, \eqs{UH}. One obtains
\begin{align}
T^{\mu\nu}-T^{\mu\nu}{}^\dagger &=  \T^\mu{}^\dagger \!  \delta \, T^\nu,  &
T_b^{\mu\nu}-T_b^{\mu\nu}{}^\dagger &=  \T_b^\mu{}^\dagger \!  \delta \, T_b^\nu.
\end{align}
 As for the hybrid case, these have the same form as usual unitarity relations, and differ from them only in that they contain $\pi N$ t matrices of different origin. The task of achieving exact 2-body unitarity for the crossing symmetric amplitudes is therefore a numerical one - the standard potential model of \eqs{t}, and its parameters, need to be adjusted so as to ensure that the  single- and double-gauged amplitudes coincide on-shell.
Finally, it is worth pointing out that 3-body unitarity could be obtained in a similar fashion by gauging a  standard Faddeev-like description of the $\pi \pi N$ system \cite{Afnan:1986sc}.

\end{document}